# Coded Path Protection
# Part 1: Efficient Conversion of Sharing to Coding

Serhat Nazim Avci *Student Member, IEEE* and Ender Ayanoglu *Fellow, IEEE*

*Abstract*—Link failures in wide area networks are common and cause significant data losses. Mesh-based protection schemes offer high capacity efficiency but they are slow and require complex signaling. Additionally, real-time reconfigurations of cross-connects threaten their transmission integrity. On the other hand, there are other schemes that are proactive. Proactivity results in higher restoration speed, lower signaling complexity, and higher transmission integrity. This paper introduces a coding-based proactive protection scheme, named Coded Path Protection (CPP). In CPP, a backup stream of the primary data is encoded with other data streams, resulting in capacity savings. In addition to a systematic approach of building valid coding structures, this paper presents an optimal and simple capacity placement and coding group formation algorithm. The algorithm converts the sharing structure of any solution of a Shared Path Protection (SPP) technique into a coding structure with minimum extra capacity. We conducted quantitative and qualitative comparisons of our technique with the SPP. Simulation results confirm that CPP provides faster link failure recovery than SPP while it incurs marginal extra capacity beyond that of SPP. In this Part 1 of the paper, we describe the theory and an algorithm for converting a given SPP solution into a CPP solution.

*Keywords*- Networks, network fault tolerance, codes, shared path protection, network coding.

## I. INTRODUCTION

Studies show that reasons of failure in networks can be widespread. According to [1], cable cut rate per 1000 sheath miles per year is 4.39. That means on average a cable cut occurs every three days per 30,000 fiber miles. These numbers are consistent with the FCC data, summarized as 13 cuts per year for every 1,000 miles of fiber and 3 cuts per year for every 1,000 miles of fiber for metro and long haul networks respectively [2]. As stated in [3], 70% percent of the unplanned network failures affect only single links. For this reason, in this paper, we focus on single link failure recovery.

Automatic protection switching (APS) in its 1+1 and 1:1 varieties were early attempts of path-based protection mechanisms. In both varieties, a dedicated spare path is employed to protect an active primary one, with the spare path being always active in 1+1 APS and activated after the failure in 1:1 APS. Neither variety was widely deployed by service providers due to their extremely low capacity efficiency [2].

Mesh-based protection schemes attracted attention due to their high capacity efficiency although they suffer from low speed. Shared Path Protection (SPP) [4–10] is a widely recognized mesh-based path protection technique. It specifies two link-disjoint paths for each connection and reroutes the traffic over the protection path if the primary path fails. It assumes up to a number of simultaneous failures and protects all possible such failures simultaneously by the sharing of protection paths.

Reference [11] introduced the concept of a $p$-cycle in order to achieve both fast restoration and low spare capacity percentage. Fundamentally, a $p$-cycle is a mixture of mesh-based protection and ring-type protection [2]. Its performance is similar to SPP in terms of resource utilization and similar to ring-type protection in terms of restoration time. Although variations exist, in its simplest form, it can be thought of a ring that goes through all the nodes in the network. In the case of a failure in a link protected by the cycle, the affected traffic is rerouted over the spare capacity in the healthy parts of the $p$-cycle. The $p$-cycle approach achieves higher restoration speed by simply minimizing the number of optical cross-connect (OXC) configurations after failure. "Hot-standby" [12] and "pre-cross-connected trials" (PXT) [13], which are extensions of SPP, are developed based on the same idea. In [14], different pre-cross-connected protection schemes are compared. The quantity and the variety of the pre-cross-connected protection schemes indicate the severity of the restoration time and stability concerns due to dynamic OXC configurations.

We offer a novel proactive protection scheme called Coded Path Protection (CPP). It is faster and more stable than rerouting-based schemes because it eliminates feedback signaling and real-time configurations after a failure. The capacity placement algorithm of CPP is based on converting the sharing operation of SPP into encoding and decoding operations with an incremental extra cost. In this Part 1 of our paper, we discuss this algorithm. Integer linear programming (ILP) is incorporated to carry out optimal conversion with minimum total capacity. ILP formulations are described, comparisons between our scheme and SPP are performed, simulations over realistic network scenarios are carried out, and their results are discussed in Part 2 of this paper.

## II. RELATED WORK

The idea of incorporating network coding into link failure protection as in this paper dates back to 1990 [15] and 1993 [16], prior to the first papers on network coding [17]. The technique is called *diversity coding*, and in its simplest form,

The authors are with the Center for Pervasive Communications and Computing, Department of Electrical Engineering and Computer Science, University of California, Irvine, CA 92697-2625, USA.

This work was partially supported by the National Science Foundation under Grant No. 0917176. Any opinions, findings, and conclusions or recommendations expressed in this material are those of the authors and do not necessarily reflect the view of the National Science Foundation.

This work was presented in part during the IEEE International Conference on Communications, Ottawa, Canada, June 2012.

$N$ primary links are protected using a separate $N + 1^{st}$ protection link which carries the modulo-2 sum, or XOR combination, of the data signals in each of the primary links. If all of the $N + 1$ links were disjoint or physically diverse, then one can recover from any single link failure by applying the modulo-2 sum over the received links. Assume that bits on the primary links are $b_1, b_2, b_3, \ldots, b_N$ and the checksum of the primary bits is

$$c_1 = b_1 \oplus\ b_2 \oplus \cdots \oplus b_N = \bigoplus_{j=1}^{N} b_j.$$

In the receiver side of the operation, if a failure is detected, the decoder applies modulo-2 sum to the rest of the $N$ links and extracts the failed bit as

$$c_1 \oplus \bigoplus_{\substack{j=1 \\ j \neq i}}^{N} b_j = b_i \oplus \bigoplus_{\substack{j=1 \\ j \neq i}}^{N}(b_j \oplus b_j) = b_i$$

where we assumed $i$ is the failed link. This operation is fundamentally different than rerouting-based protection schemes since it does not need any feedback signaling. In this paper, for simplicity, we will use XOR and regular summation notations interchangeably, in the expectation that the meaning will be clear from the context.

This idea was revisited by the authors of this paper in [18, 19] and a coding structure for an arbitrary network topology was developed. This scheme may require extra links from the destination nodes to decoding nodes to be able to decode signals. It has been shown in [18] that diversity coding can achieve higher capacity efficiency than the SPP and the $p$-cycle techniques in some networks. In the same paper, diversity coding has been shown to perform better than the other two when both the capacity efficiency and restoration speed are jointly taken into account. Optimal design algorithms for diversity coding are developed in [20], for both pre-provisioning of the static traffic and the dynamic provisioning of the dynamic traffic. In [21], it is shown that with proper buffering and synchronization, diversity coding can achieve sub-ms restoration time. In [22], the basic structure of diversity coding is extended to incorporate both primary and protection paths in coding operations, resulting in improvement in capacity efficiency. The idea of converting an SPP solution to a coding-based solution is introduced by the authors in [23]. Preliminary results from suboptimal simulations have validated the potential of this idea.

In [24], a bidirectional protection scheme that uses network coding over $p$-cycle topologies on mesh networks was introduced and called as 1+N protection. The idea presented in [24] is to form circular protection paths in both directions that traverse the source and destination nodes of the group of flows that are to be protected. In [25], a new tree-based protection scheme was introduced instead of a $p$-cycle based scheme and called Generalized 1+N protection (G1+N). In [25], same data from both end nodes are sent on a parity link. Symmetric transmission is broken only for the connection affected from the failure. The capacity efficiency of G1+N is not available in the literature. However, it clearly lacks the speed of diversity coding due to two different delays. First, the distance between the decoding node and the destination nodes are always higher than the half of the propagation delay over the coding tree. Second, the backup stream is transmitted only after a certain delay until the primary stream, transmitted over the primary path, is received by the destination node. In [26], a new trail-based protection scheme is proposed. Failed data is recovered via a linear coded protection circuit. This structure is a modified version of the scheme in [24] resulting in higher capacity efficiency by moving from cyclic to linear protection topology.

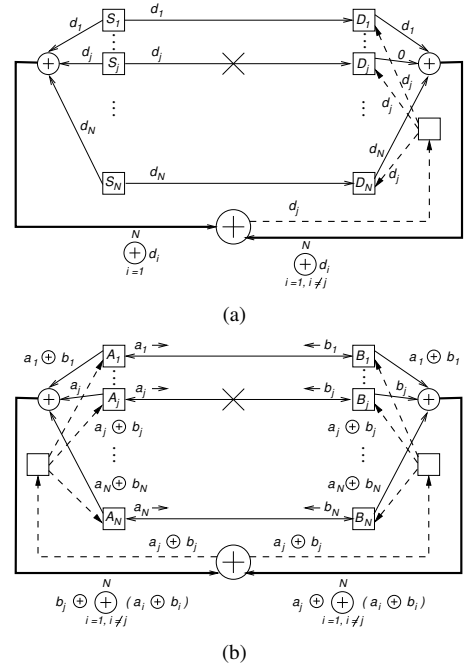

Fig. 1. Multipoint-to-multipoint architectures with coded protection against single link failures. (a) Unidirectional network [15, 16], and (b) bidirectional network [24, 25].

Protection of $N$ parallel unidirectional links via coding is shown in Fig. 1(a) [15, 16], and $N$ parallel bidirectional links in Fig. 1(b) [24, 25]. Although it has more demanding synchronization requirements, in this paper, we studied conversion of a bidirectional SPP solution for a network with arbitrary topology to a bidirectional CPP solution for it. We plan to pursue a study of unidirectional networks next.

## III. Coded Path Protection

In this paper, we propose a novel coding technique, which we call Coded Path Protection (CPP). We present a simple strategy to find the optimal coding structure without much complexity. CPP is faster, has less signaling complexity, and has higher transmission integrity than any of the rerouting-based protection techniques. On the other hand, spare capacity percentage (SCaP) of CPP is slightly larger than the SCaP of SPP. Our contribution in this paper consists of two parts, namely a novel coding structure and a simple but optimal coding group formation algorithm.

Our technique is applicable to networks whose links are bidirectional or unidirectional. In this paper, we investigate networks with bidirectional links and symmetric traffic on each link. In real networks, increasingly, traffic in the two directions of a link may be asymmetric. We can model such links with two unidirectional links of different transmission rates. We leave a study and investigation of the tradeoffs of restoration speed versus SCaP for networks whose links can be decomposed into two unidirectional ones, regardless of symmetric or asymmetric traffic, for future study. In a similar fashion, our technique is applicable to networks where transmission is electrical or optical, and encoding and decoding operations can be performed in the electrical or optical domain. Encoding and decoding for single link failures can be accomplished by using XOR logic. As discussed later, currently, this can be done in the optical domain. For the purposes of this paper, we assume optical transmission and cross-connects, with encoding and decoding performed in the optical domain. Again, we leave a study of the options for electrical processing for future study.

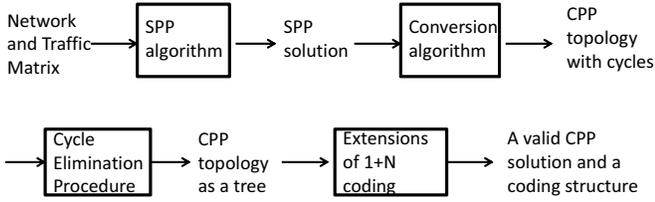

Fig. 2. Steps in creating a CPP solution.

In Fig. 2, we illustrate how the input network and traffic data are processed in order to reach a CPP solution with a valid coding structure, in sequential order. In this section, we begin its description by explaining our methodology in converting a typical solution of SPP into one with sharing replaced by coding. We show how to establish a valid coding structure and show that the encoding and decoding inside the network can be carried out within this coding structure. In the next section, first, we explain a design algorithm that finds an SPP solution. Second, we present a design algorithm, which optimally converts the sharing structure into the coding structure given the solution of SPP.

We benefit from the basic coding structure of [26] while building a valid coding structure in CPP. When the traffic is bidirectional, the end nodes of a connection generate the same set of protection signals and transmit them over the protection path to the other end node of the connection so that the failed data can be recovered from the protection path shortly after the failure. This proactive protection mechanism makes 1+N coding [26] faster than the sharing-based protection schemes. This structure creates a symmetry over the protection path of each bidirectional connection. The parity data is formed by applying the XOR operation to the data received from and the data transmitted over the primary path. Encoding and decoding of different parity data inside the network can be carried out by utilizing the symmetry over the protection paths of the connections. Despite its restoration time advantage, 1+N coding works in a specific limited linear topology, which is why it falls short in exploring the full connectivity inside the network. On the other hand, CPP mostly preserves the topology of an SPP solution with a small compromise on connectivity without losing the speed advantage that is inherent in the coding structure.

Symmetric transmission is key in the encoding and decoding operations. This is illustrated in Fig. 3(a) in an example with two connections. Thick straight lines are primary paths and dotted lines are protection paths. For the time being, synchronization and timing are not considered. Assume that $S_1$ transmits $s_1$ to $D_1$ and $D_1$ transmits $d_1$ to $S_1$ using the primary path at time $t_0$. After a delay of $\tau$, these signals are received by the reciprocal nodes at the same time and both end nodes form the summation of these two signals, mathematically $c_1 = s_1 \oplus d_1$. At time $t_0 + \tau$, the same $c_1$ symbols are sent from the corresponding end nodes of the protection path of $S_1 - D_1$. It is similar for $S_2 - D_2$. As it is seen at Fig. 3(a), $c_1$ and $c_2$ are coded over the link $A - B$. $A$ is the node where $c_1$ and $c_2$ are coded and node $B$ is responsible from decoding (node $A$ in the opposite direction): Node $B$ extracts $c_1$ using $c_1 \oplus c_2$ and $c_2$, and extracts $c_2$ using $c_1 \oplus c_2$ and $c_1$. Therefore encoding and decoding of signals $c_1$ and $c_2$ are successfully completed and node $B$ transmits them to $D_1$ and $D_2$, over links $B - D_1$ and $B - D_2$ respectively. Fig. 3(b) gives an example of single link failure on the primary path of $S_1 - D_1$. Due to the failure, $S_1$ receives 0 instead of $d_1$ and transmits $s_1 \oplus 0 = s_1$ at time $t_0 + \tau$ over the protection path. Similarly, $D_1$ receives 0 instead of $s_1$, so it transmits $0 \oplus d_1 = d_1$ over the protection path. These signals are coded with $c_2$ over link $A - B$ and they are decoded at nodes $B$ and $A$ respectively. At node $B$, $s_1$ is extracted by summing $s_1 \oplus c_2 \oplus c_2 = s_1$ and forwarded over the link $B - D_1$. At node $A$, $d_1$ is acquired by summing $d_1 \oplus c_2 \oplus c_2 = d_1$ and forwarded over the link $A - S_1$. It can be observed that other single link failures over the primary paths will result in recovery of the failed link over their corresponding protection paths.

"Poison-antidote" analogy [27] is useful in understanding the general coding structure. When two signals are coded together, they "poison" each other. At the decoding node, "antidote" data are needed to extract the signals from each other. For the general two connection case in Fig. 4, same signals traverse the reverse directions over the protection path. Straight lines are the protection paths of $a$ and $b$ in one direction. Dotted lines are the protection paths of same $a$ and $b$ in the reverse direction. At node $A$, straight paths are antidotes of dotted paths. At node $B$, dotted paths are antidotes of straight paths. In the single link failure case, if connections have link-disjoint primary paths, at most one of them is affected from the failure. The other connection can preserve reciprocity and the poison-antidote structure to help recover the affected connection. Note that the term "link-disjointness" actually refers to span-disjointness. We will employ the former term since it is used more commonly in the literature.

We can generalize this coding structure to an arbitrary number of connections, arbitrary number of links, and to an arbitrary topology by the use of reciprocity. However, first we should define our concept of a coding group. Assume that $S_1 - D_1$ and $S_2 - D_2$ are coded over some link $E$. Then, they are considered to be in the same coding group. This group gets bigger if either $S_1 - D_1$ or $S_2 - D_2$ is coded with some other connection. For example, if $S_3 - D_3$ is coded with $S_2 - D_2$, it

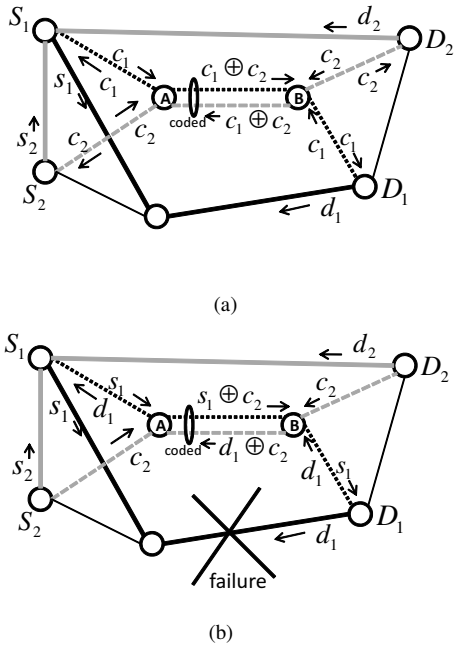

(a)

(b)

Fig. 3. Coding and decoding operations for coded path protection, (a) In normal state, (b) Reaction to single link failure

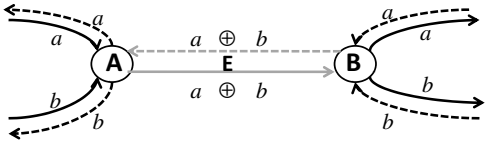

Fig. 4. Coding at an arbitrary link and decoding at an arbitrary node.

is also considered in the same coding group with both $S_2 - D_2$ and $S_1 - D_1$. In a coding group, coding structure can recover from a single link failure on one of the primary paths if the reciprocity property is preserved for the other connections. To guarantee this property, two protection paths can be coded together as long as

1) Their primary paths are link-disjoint,
2) Their primary paths are also link-disjoint with the primary and protection paths of the connections in the same coding group.

These are sufficient link-disjointness rules to satisfy the decodability on arbitrary CPP topologies. In Section III-D, these rules are relaxed to some extent to utilize the network connectivity more. These rules can also be interpreted as the criteria of two connections to be in the same coding group. In addition to these rules, the primary and protection paths of the same connections are inherently link-disjoint as a very significant rule in path-based link failure recovery. CPP is a generalization of the scheme in [26], with two differences. First, CPP is for an *arbitrary network topology,* whereas the scheme in [26] has the fixed structure of a number of disjoint links. Second, the capacity efficiency of CPP is quantified in this paper, while for the scheme in [26], it is unknown. CPP is suitable to convert a typical solution of SPP with low complexity because a typical solution of SPP must obey the first rule above. The rest of the work to convert an SPP solution

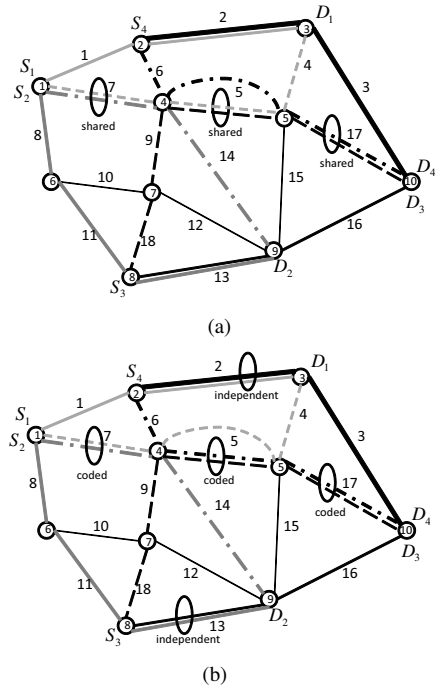

Fig. 5. Possible coding and sharing scenarios over a network, (a) Sharing of protection capacities, (b) Coding of protection paths.

to CPP is to form coding groups that satisfy the second rule.

### A. Conversion Example

We assume that for a given topology and a given set of connections, there is a pre-calculated solution of SPP. Given the solution, primary and protection paths of the connections, wavelength assignments, and maximum required spare capacity on each link will be known. Referring to Fig. 5(a), thick straight lines represent the primary paths of end-to end connections, whereas protection paths are shown by dotted lines. In Fig. 5(a), numbers associated with the edges are their index values. Some of the protection capacity is shared by multiple protection paths. There is limited freedom in terms of choosing the group of connections which will share the same capacity over the same link. For example, $S_3 - D_3$ can share the one unit spare capacity at link 5 either with connection $S_1 - D_1$ or with connection $S_4 - D_4$. However, $S_1 - D_1$ and $S_4 - D_4$ cannot share that capacity since their primary paths are not link-disjoint. This freedom can be utilized in converting sharing groups to valid coding groups with zero or unappreciable additional capacity.

In the given solution of SPP, protection paths are coupled under the provision of the first rule. However, while building the CPP solution, protection paths are coupled and coding groups are formed in a way such that both rules are satisfied. The sharing structure in Fig. 5(a) is converted to the coding structure in Fig. 5(b) in this manner. It should be noted that at link 5 $S_1 - D_1$ and $S_3 - D_3$ are coupled to share the one unit capacity in the SPP solution. However, in the CPP solution $S_3 - D_3$ is coded with $S_4 - D_4$, not with $S_1 - D_1$. If that is not done, then $S_1 - D_1$, $S_2 - D_2$, $S_3 - D_3$ and $S_4 - D_4$ would be in the same coding group because they would be indirectly related. Then the second rule about link-disjointness would not be satisfied. After this modification, we can divide this coding group into two, one group consists of $S_1 - D_1$ and $S_2 - D_2$ and the other consists of $S_3 - D_3$ and $S_4 - D_4$. Then both of the rules are satisfied. In this example, no extra capacity is required to convert an SPP solution to a CPP solution with the aid of limited freedom in the SPP solution. However, that is not the case in general. Therefore, in Part 2 of this paper, we describe an ILP formulation to conduct the conversion with minimum extra capacity.

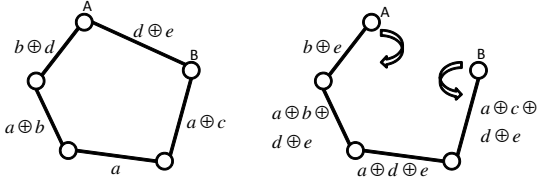

Fig. 6. Removal of a cyclic structure.

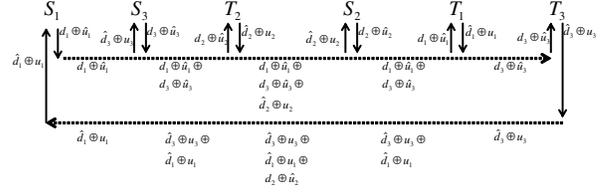

Fig. 7. Coding at 1+N protection circuit for 3 connections.

### B. Cycle Elimination

The outputs of the conversion algorithm are the coding group combinations and their protection topologies. A valid encoding and decoding structure will be built upon the assumption that the protection topology has a tree structure which means it includes no cycles. It is known that cyclic structures inside coding topologies can impair the encoding and decoding structures. In addition, cyclic structures are less capacity efficient than tree structures. It is stated in [25], in its Proposition 1, that "*under the assumption of undirected edges in the network graph G, the minimal cost protection circuit, $P_i$, where the cost is in terms of the number of network edges, is a tree.*" There are two ways to eliminate those cyclic structures to transform the coding group topologies into tree structures. First, the conversion algorithm can be modified to prevent those cyclic structures from occurring. However, this method significantly increases the complexity of the conversion algorithm of CPP. Second, a handful of cyclic structures can be eliminated by the conversion algorithm via a simple method, which we call "cycle elimination procedure" (CEP). As a side advantage, CEP results in further capacity savings. The CEP is shown in Fig. 6 with an example. The network in the figure is a portion of the coding topology. The protection data of five different connections are coded in both directions as shown in the Fig. 6. The data on the longest link, which is the link between node $A$ and node $B$, of this cyclic structure can be rerouted and coded with the data over the rest of the cycle. The link $A - B$ is emptied and the data $d \oplus e$ are coded with the data over the rest of the cycle. It eliminates the cyclic property of this portion of the coding topology (which ensures the tree structure) and results in saving of the capacity of the longest link. The link-disjointness rules ensure that the rest of the cycle does not share any link with the primary paths of the connections of interest.

### C. CPP Coding Structure

We need to prove that simple linear coding structure of 1+N coding can be extended to any arbitrary tree structure in order to implement this idea over arbitrary CPP protection topologies, which consist of trees. Before demonstrating how to build a general coding structure for CPP, we need to show the extensions that can be built over the simple linear coding structure of 1+N coding. The basic structure of 1+N coding protection circuit (trail) for 3 connections is shown in Fig. 7. For clarity, the link-disjoint primary paths between the end nodes are not shown. The symbols with the hat sign in the parity data are the signals that are received from the primary paths. For an $S_i - T_i$ end node pair, the parity data generated at both end nodes are the same except the hat signs are on top of different signals as in $d_i \oplus \hat{u}_i$ and $\hat{d}_i \oplus u_i$. For any node on the trail, in no failure state, the input and output signals are the same except the position of the hat signs. The symbols with the hat sign complement each other in the corresponding parity signal. If one of the primary paths fails, the corresponding symbols in the parity data with the hat sign becomes zero. If the primary path of $S_i - T_i$ pair fails, then the parity signals of $S_i - T_i$ become $d_i$ and $u_i$, respectively.

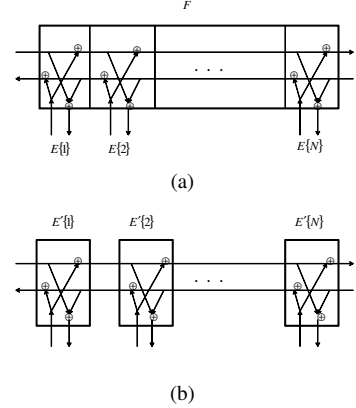

Fig. 8. Proof of *Lemma 1* (a) Multiple end nodes share the same node, (b) Each end node can be shown as a separate entity over the protection trail.

*1) Lemma 1: In the linear topology of 1+N coding, a node can serve as the end nodes of multiple connections.*

The proof is intuitive as shown in Fig. 8. In this case, these end nodes can be represented by separate hypothetical adjacent end nodes on the linear coding graph (trail). The links between these end nodes are assumed to have zero length. In other words, multiple end nodes over the linear coding structure may refer to the same physical node if the links between them has zero length. Each end node can be separated from each other since they are connected to the physical node via independent ports as shown in Fig. 8(a). The parallel horizontal links represent the coding trail passing through the nodes of interest. Let $E$ be the set of end nodes which share the same physical node $F$. Since the information regarding each end node is independent of the other nodes, they can be separately depicted with the hypothetical end nodes in Fig. 8(b).

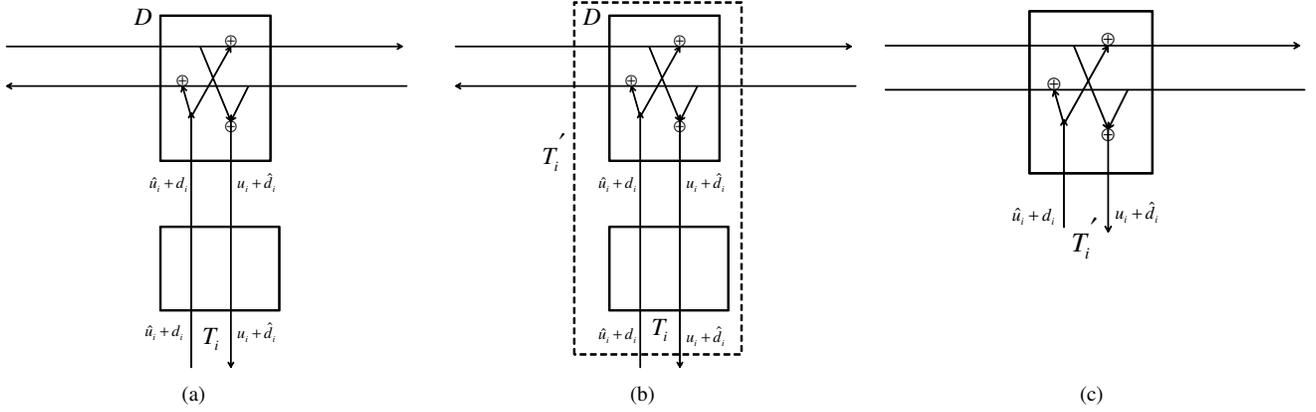

Fig. 9. Proof of *Lemma 2* (a) An end node is connected to the coding trail via a direct path, (b) From trail point of view, they are seen as a single node, (c) The end node hypothetically is over the trail.

*2) Lemma 2:* The classical 1+N coding requires each end node to be traversed by the common protection path. However, the same coding structure can be applied even if an end node is connected to the linear topology through a direct path which deviates from the common trail.

Refer to Fig. 9 for the proof. In Fig. 9(a), the end node $T_i$ is connected to the linear coding topology via an arbitrary on-trail node $D$ through a bidirectional link. From the coding trail perspective at node $D$, there is no change if the node $D$ and node $T_i$ are merged into a single hypothetical node $T_i'$. This transformation is depicted in Fig. 9(b), where the dashed box combines these two nodes into a single one in terms of coding operations over the linear topology. In Fig. 9(c), Fig. 9(b) is simplified and node $T_i$ is represented on the linear coding trail via a hypothetical node $T_i'$. We can generalize this operation to any arbitrary number of end nodes as long as they are connected to the common trail via link-disjoint paths.

*3) Lemma 3:* As an extension to the Lemma 2, if $N$ end nodes are connected to a node on the trail via a common link, these end nodes can be still represented over the trail by a different notation.

In Fig. 10(a), end nodes which are in the set of $K$ are combined at an arbitrary node $C$, and $C$ is connected to an arbitrary node $D$ over the linear coding trail. In this case, the end nodes $S_i \in K$ and $T_j \in K$ cannot be represented independently because there is no mechanism to decode the signals in node $C$. However, from the network point of view, these multiple end nodes can be merged into a single end node as

$$\sum_{i, S_i \in K} S_i' + \sum_{j, T_j \in K} T_j'$$

since the coding operations in the rest of the network are not affected. Note that $\sum_{i, S_i \in K} S_i' + \sum_{j, T_j \in K} T_j'$ is only a notation because the end nodes cannot be summed but their parity signals are summed. The new hypothetical node is depicted in Fig. 10(b). In Fig. 10(c), the node $\sum_{i, S_i \in K} S_i' + \sum_{j, T_j \in K} T_j'$ is hypothetically placed over the trail using *Lemma 2*. The number of combined end nodes can be set to an arbitrary number $N$ and the hypothetical end node will be represented as the summation of all the combined end nodes. This *lemma* is useful if the separate signals of these end nodes are not of interest.

*4) Lemma 4:* If we merge any arbitrary number of adjacent end nodes over the linear coding trail, the coding operations in the rest of the trail are unaffected.

Refer to Fig. 11 for the proof. Let $P$ be the set of adjacent end nodes which are supposed to be merged into a single end node over the trail. In Fig. 11(a), the encoding and decoding operations inside these end nodes are shown. This structure can be converted to the structure in Fig. 11(b), if the individual signals of the end nodes in $P$ are not necessarily extractable. Then, the combination of these end nodes is represented with a single hypothetical end node as shown in Fig. 11(c).

*5) Lemma 5:* If the extensions to the linear coding trail do not create a cyclic structure inside the topology, it is possible to separate the topology into two subsystems.

Similarly to the previous *lemmas*, the proof is intuitive as shown in Fig. 12. In Fig. 12(a), the coding topology can be divided into two subsystems with the dashed link. These subsystems are highlighted in Fig. 12(b). One of them is the hypothetical end node, defined by *Lemma 3*, represented as the summations of multiple end nodes, which are spanned by a common link from the linear coding trail. The other subsystem is the rest of the coding topology, which is the rest of the tree. Regarding the input and output relationship between these two subsystems at that specific branch point $D$, it is seen that these subsystems are the complements of each other. The complementary hypothetical end node is formulated in Fig. 12(b).

The fact that encoding and decoding operations can be carried out at the hypothetical end nodes and the branch points accomplish the sought-after and elusive implementation by the network coding community of encoding and decoding inside the network for multiple unicasts.

*6) Example 1:* In order to visualize how these lemmas are useful in transforming a tree topology into a linear trail

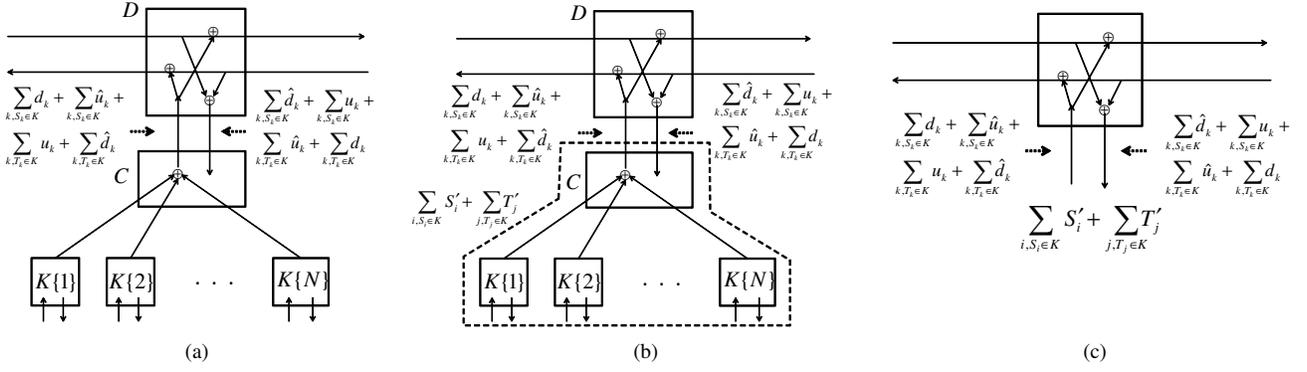

Fig. 10. Proof of *Lemma 3* (a) Two different end nodes are connected to the trail via the same link, (b) They can be merged into a single node, (c) How they are seen from the rest of the trail.

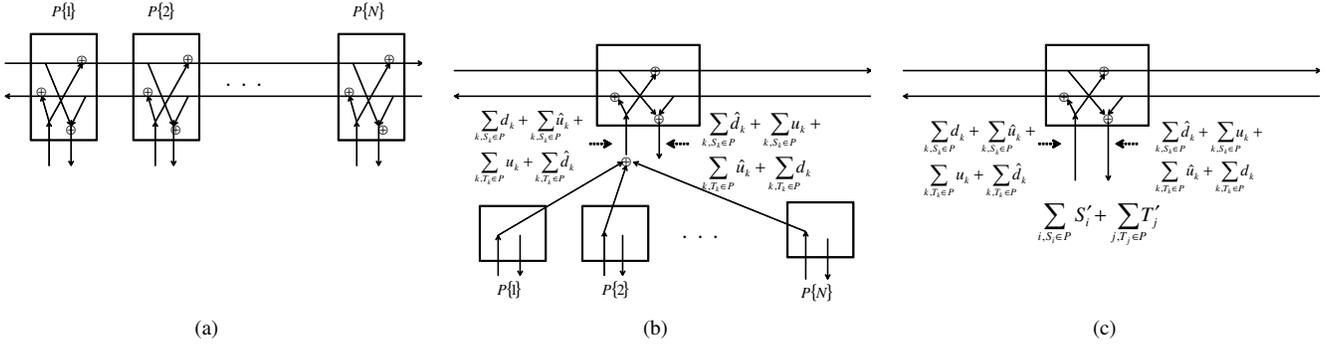

Fig. 11. Proof of *Lemma 4* (a) Multiple adjacent end nodes over the trail are depicted, (b) These end nodes can be merged into a single one, (c) How they are seen from the rest of the trail.

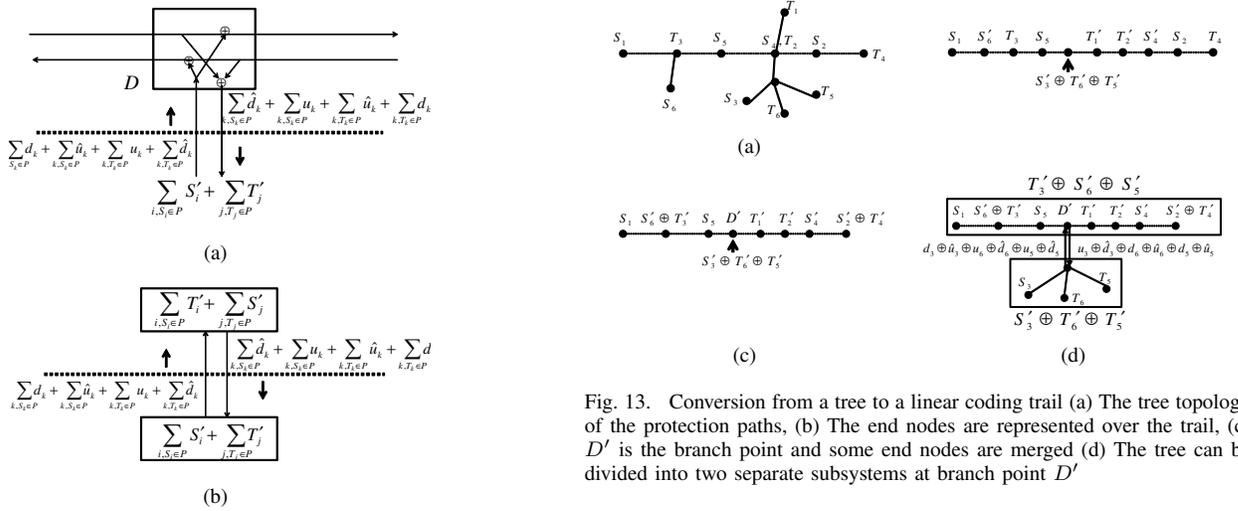

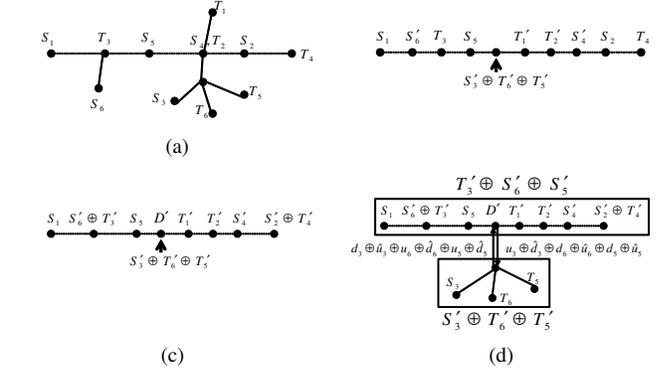

Fig. 12. Proof of *Lemma 5* (a) Input-output relationships between the end node and the rest of the coding trail, (b) The rest of the trail is treated as the combination of some of the end nodes

Fig. 13. Conversion from a tree to a linear coding trail (a) The tree topology of the protection paths, (b) The end nodes are represented over the trail, (c) $D'$ is the branch point and some end nodes are merged (d) The tree can be divided into two separate subsystems at branch point $D'$

topology, an example is provided below. Assume that, there are 6 bidirectional connections such that $S_i$ is communicating with $T_i$ for $i = 1, 2, 3, 4, 5, 6$. There exists a bidirectional primary path between each end node pair which is link-disjoint with the other primary paths and with the common protection trail. In Fig. 13(a), the end nodes of the connections are shown on the network. For clarity, link-disjoint primary paths are not depicted. The dashed link is the linear coding trail which has the coding structure of 1+N coding. The protection paths have a topology with a tree structure. Using *Lemmas 1 to 3*, we can convert this tree topology into a trail coding topology. The hypothetical nodes are highlighted with a prime sign. They are no different than regular end nodes in terms of coding operations over the trail. The converted structure is given in Fig. 13(b). The end nodes, which are shown as single entities over the trail, can successfully extract their parity data from the trail. In the case of a failure in their primary paths, they can recover the failed data from the trail as shown in [26].

In the next step, some of the adjacent end nodes are merged. In Fig. 13(c), the end node pairs $S_2 - T_4$ and $T_3 - S_6$ are merged into single hypothetical nodes using *Lemma 4*. In Fig. 13(d), at the specific branch point $D$ over the trail, the coding topology is divided into two subsystems. The underlying topologies are shown inside the boxes. The notations outside the boxes are the images of the subsystems as they are seen from the opposite subsystem.

*7) Coding Strategy:* Assume that there is a coding group consisting of $N$ connections, which are given in the set $P = \{S_i, T_i : 1 \leq i \leq N\}$ meaning that each connection consists of the end nodes with the same indices. As stated before, the protection topology of this coding group is link-disjoint with the primary paths in the same coding group and it is a tree. The end nodes of the connections are scattered over this tree. A valid encoding and decoding structure is established using the following steps.

1) Select one of the links inside the tree and call it the truck trail.

2) Extend this truck trail from both ends as long as the extended links reach at the edges of the tree. When there are multiple links to extend, one of them is randomly selected.

3) When the truck trail reaches its limits, using *Lemmas 1 to 3*, place the end nodes over the trail. There are three types of end nodes. The end nodes which are physically over the trail are shown as separate entities over the trail with help of *Lemma 1*. The second type of end nodes are not physically over the trail but directly connected to the trail via a dedicated path. They are depicted over the trail with the help of *Lemma 2*. The third type of end nodes are connected to the trail via a common link or common links. These end nodes are placed over the trail as a combination of multiple end nodes with the help of *Lemma 3*. We call the hypothetical nodes which represent the combination of multiple end nodes as the branch points on the trail. There can be multiple branch points over a single trail.

4) Assume $R$ is the set of combination of multiple end nodes as $R = \{R_1, R_2, ..., R_k\}$, where $k$ is the total number of branch points over the truck trail. Each $R_i$ keeps the end nodes that are spanned by the branch point $i$. If the same pair of end nodes belong to the same set $R_i$, $1 \leq i \leq k$, they are omitted from the truck trail. They will be taken into account later.

5) Then, code the signals whose end nodes are over this truck trail as it is explained in [26] under 1+N protection coding operations. The truck trail is the protection circuit of 1+N coding. The end nodes which are shown as single entities will be able to receive their parity data from the trail. In the case of failure, these nodes are able to extract the failed data from the linear 1+N coding trail.

6) The remaining end nodes are the ones that are depicted as the combination of multiple end nodes. There are $k$ combinations and each combination has a branch point. Originating from these branch points, new branch trails will be initiated using the links that span the end nodes in sets $R_i$, $1 \leq i \leq k$.

7) Consider the set of $R_1$ and the branch point of this set. Include the end nodes that are omitted from this set at step 4. We initiate a branch trail originating form the branch point of this set. The link that connects the end nodes in $R_1$ to the truck trail is the first link of this branch trail. Extend this branch trail to the opposite direction of the branch point as long as the trail reaches the edge of the branch. When there are multiple options, randomly pick one of the links to extend the branch trail.

8) Using *Lemma 5*, we can define the branch point as the starting point of this trail. This point behaves as the complement of the end nodes combined at this branch point. For example, if the combined end nodes are $S_i \oplus T_j$, then the branch point would be seen as $T_i \oplus S_j$ over the branch trail.

9) Place the end nodes over this trail using *Lemmas 1 to 4* as in step 3.

10) Repeat step 4 and 5. $R_1 = \{R_{1,1}, R_{1,2}, ..., R_{1,k_1}\}$, where $k_1$ is the number branch points over the first branch trail.

11) Return to step 6 iteratively as long as all of the sub-branches of the first branch are explored and each end node is placed over a branch trail as a single entity. This will make sure that every end node spanned by the first branch point is able to receive its parity data form the tree.

12) Pass to the next branch over the truck trail and return to step 7.

At the end, all of the end nodes in CPP tree topology will be placed as a single entity in one of the linear 1+N coding trails, which makes them protected against single link failures. To clarify the steps shown above, an example is provided below.

*8) Example 2:* In Example 1, a tree structure was partially converted to a linear 1+N coding trail. This trail is the truck trail to start with. We proceed from Example 1 with an additional connection demand between $S_7$ and $T_7$. The updated coding group topology is shown in Fig. 14(a). In Fig. 14(b), it is shown that all of the end nodes except $S_3'$, $T_6'$, $T_5'$, $S_7'$, and $T_7'$ are placed over the linear 1+N coding trail, which enables them to encode and decode their parity data over this trail. As shown in Fig. 14(a), the protection path between $S_7$ and $T_7$ is physically separated from the truck trail which is shown with dashed links. That means the signals of $S_7 - T_7$ are bounded within the branch originating from node $D'$ of the truck trail. Therefore, in Fig. 14(c), we omit the end nodes $S_7'$ and $T_7'$ from the truck trail as explained in step 4 in Section III-C7.

In order to protect the end nodes which are not singularly shown over the truck trail, we need a branch trail originating from the branch point $D'$. As shown in Fig. 14(d), the branch point is considered as $T_3' \oplus S_6' \oplus S_5'$ replacing the rest of the trail. According to *Lemma 4*, there is no need to show $T_3'$, $S_6'$, and $S_5'$ as separate end nodes over the branch trail. As explained in step 7, we reincorporate the connection $S_7 - T_7$ for the branch trail since its protection path resides over the branch trail this time. The branch trail is extended as defined in step 2 in the previous section. The end nodes that are spanned by this branch are placed over a new linear coding trail. This trail is depicted in Fig. 14(e). The reintroduction of the connection $S_7 - T_7$ does not affect the coding operations in the rest of the network since the input and output signals at the branch point are the same. The end nodes that are shown as single entities are protected by this coding trail. As in the truck trail, there is a branch point $E'$ that combines multiple end nodes over the branch trail. It is required to go one more level down and generate a sub-branch trail to cover these end nodes as well. This sub-branch trail is shown in Fig. 14(f). The operation is stopped when all of the end nodes are placed over a linear 1+N coding trail.

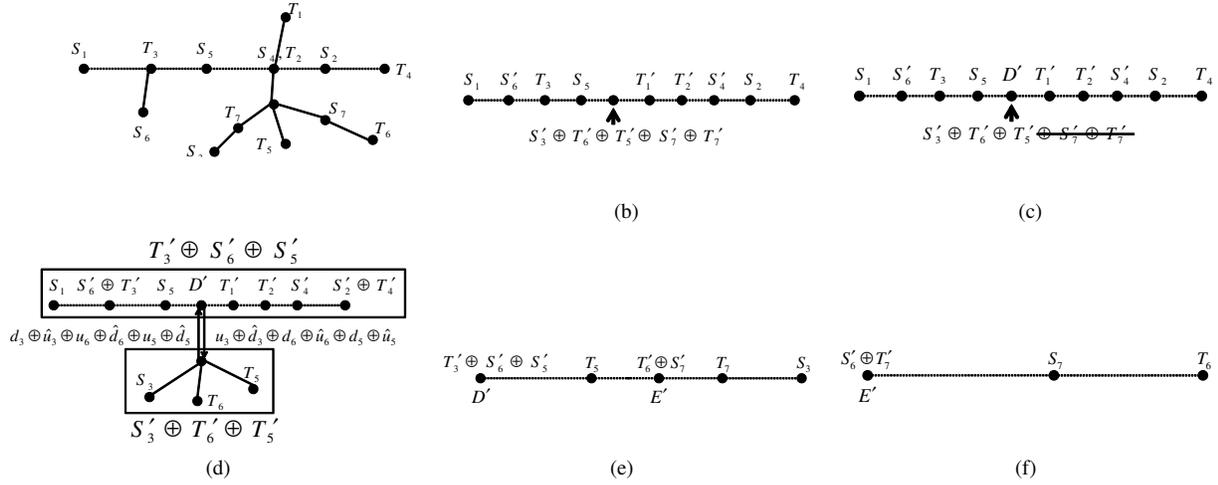

Fig. 14. Protection of each connection by creating linear branch trails (a) The updated tree topology of the protection paths, (b) The end nodes are represented over the trail, (c) $S_7$ and $T_7$ are omitted from the truck trail since they are bounded within the branch point $D'$ (d) The tree can be divided into two separate subsystems, (e) The branch trail originating from node $D'$ and connection pair $S_7 - D_7$ is reincorporated, (f) The sub-branch trail originating from node $E'$ of the branch trail.

### D. Extensions on the Coding Group Selections

The link-disjointness rules that guarantee the decodability of CPP structures are sufficient rules. These rules can be overridden within some limits without impairing the decodability of the coding structure. Therefore, there is still room to improve in terms of capacity efficiency. The first rule of link-disjointness is a necessary condition for decodability. However, the second rule can be modified to allow sharing of a common link by the primary path of a connection and the protection paths of other connections in the same coding group. The second rule is altered as

- Their primary paths are also link-disjoint with the primary paths of the connections in the same coding group.

In this mode of operation, if the common link shared by one primary path and one or more protection paths fail, then the end nodes of the failed protection paths can detect the failure over their protection paths and temporarily terminate transmission over their protection paths. Otherwise, symmetry is broken for more than one connection on the protection topology and the decoding structure crushes. In other words, the failed protection paths need to stop poisoning other protection paths because there are no antidotes. Failure detection can be carried out by comparing the data received from the primary paths with the data received from the protection paths.

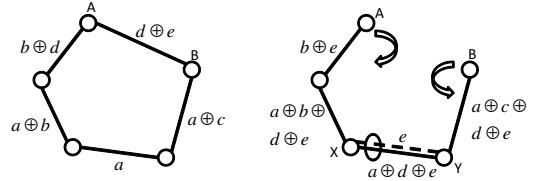

Fig. 15. Overlap of the primary and protection paths in CEP.

CEP for this mode of operation is not as straightforward as it was in the previous case. Previously, the data over the longest link of a cycle was coded with the data over the rest of the cycle and that link was released from the coding topology. However, that may not be possible when the primary paths and protection paths of different connections share a common link. In that case, the protection path of a connection can overlap with its own primary path if it is rerouted and coded over the rest of the cycle as depicted in Fig. 15. Previously, the primary paths and the protection topology were link-disjoint. However, in this mode they can share some links. After the protection data $e$ is rerouted and coded over the rest of the cycle, it can overlap with its primary data $e$ over link $X - Y$, which makes the recovery impossible if that common link ($X - Y$) fails.

To preserve the link-disjointness criterion between the primary and protection paths of the same connection, a new CEP is proposed.

1) Select the longest link on the cycle. Remove this link and code the data on it with the data over the rest of the cycle. Check if this breaks down the link-disjointness between the primary and protection paths of each connection.
2) If so, select the next longest link until a link whose removal does not affect the link-disjointness criterion is found.
3) If there is no such link, look for a separation point on the cycle. A separation point on the cycle is a node whose incoming (on-cycle) links carry no mutual data. In other words, this node is the end node of the data on both of its incoming links. If there is such a separation point, this cyclic structure can be considered as a tree structure and it preserves the coding structure.

4) If there is no separation point on the cycle, then reroute the portions of the protection paths that cause the conflict between link-disjointness and cyclic property.

5) If no solution is found then remove the connections which cause the conflict from the coding group. Protect these connections by 1+1 APS.

## IV. CONCLUSION

In this paper, we introduced a proactive network restoration technique we call Coded Path Protection (CPP). The technique makes use of symmetric transmission over protection paths and link-disjointness among the connections in the same coding group. We modified the coding structure and leveraged its flexibility to convert the sharing structure of a typical solution of SPP into a coding structure of CPP in a simple manner. With this approach, it is possible to quickly achieve optimal solutions. As a result of this operation, the CPP algorithm achieves significantly faster restoration. In this Part 1 of our paper, we described an efficient conversion algorithm from a given SPP solution to our CPP solution. In Part 2 of this paper, we will discuss an integer linear program for implementing CPP, the issues of synchronization and others that affect networks due to the introduction of CPP, as well as the results of simulations with a number of networks from the literature.